\documentclass[%
reprint,
showpacs,preprintnumbers,
 amsmath,amssymb,
 aps,
longbibliography,
floatfix,
]{revtex4-1}

\usepackage{graphicx}
\usepackage{dcolumn}
\usepackage{bm}

\begin{document}

\preprint{APS/123-QED}

\title{Alignment-to-orientation conversion in a magnetic field at nonlinear excitation of the $D_2$ line of rubidium: experiment and theory}

\author{M. Auzinsh}%
 \email{marcis.auzins@lu.lv} 
 \author{A. Berzins}
\author{R. Ferber}
\author{F. Gahbauer}
\author{L. Kalvans}
\author{A. Mozers}
\author{A. Spiss}%
 \affiliation{%
 Laser Centre, University of Latvia, Rainis blvd. 19, LV-1586 Riga, Latvia
}%

\date{\today}

\begin{abstract}
We studied alignment-to-orientation conversion caused by excited-state level crossings in a nonzero magnetic field of both atomic rubidium isotopes.
Experimental measurements were performed on the transitions of the $D_2$ line of rubidium.
These measured signals were described by a theoretical model that takes into account all neighboring hyperfine transitions, the mixing of magnetic sublevels in an external magnetic field, the coherence properties of the exciting laser radiation, and the Doppler effect. In the experiments laser induced fluorescence (LIF) components were observed at linearly polarized excitation and their difference was taken afterwards. By observing the two oppositely circularly polarized components we were able to see structures not visible in the difference graphs, which yields deeper insight into the processes responsible for these signals. We studied how these signals are dependent on laser power density and how they are affected when the exciting laser is tuned to different hyperfine transitions. The comparison between experiment and theory was carried out fulfilling the nonlinear absorption conditions.
\end{abstract}


\maketitle

\section{\label{sec:level1}Introduction}
\begin{figure}[b]
	\includegraphics[width=\linewidth]{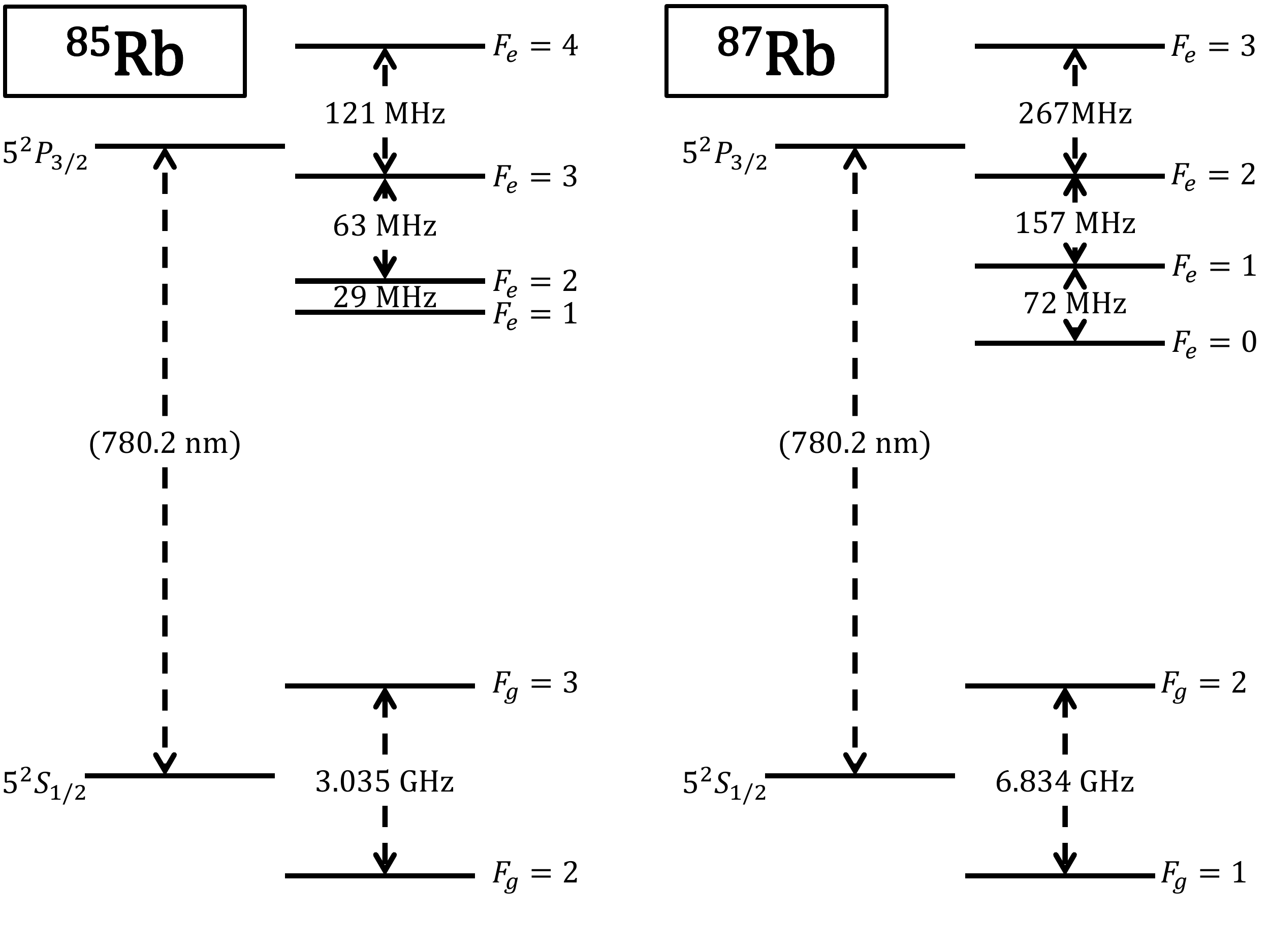}
    \caption{\label{fig:im2}Fine and hyperfine energy-level splittings for the $D_2$ transitions of $^{85}$Rb and $^{87}$Rb.}
\end{figure}
\begin{figure}[b]
	\includegraphics[width=\linewidth]{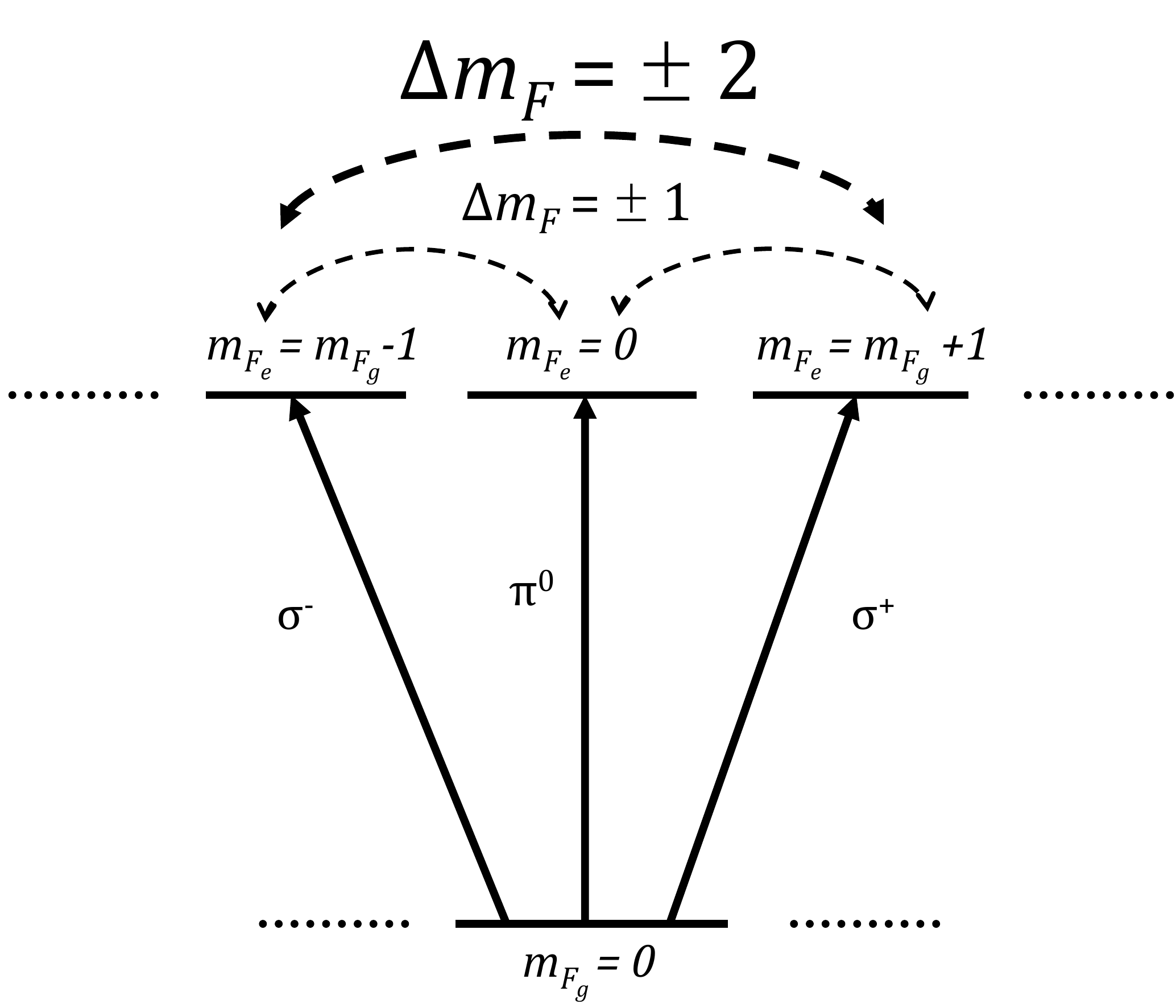}
    \caption{\label{fig:im1}Absorption from the ground-state hyperfine magnetic sublevel $m_{F_g}$ 
    and creation of $\Delta{m_F}$=1 and $\Delta{m_F}$=2 coherences in the excited state when the magnetic field $B$ = 0.}
\end{figure}
\begin{figure}[b]
	\includegraphics[width=\linewidth]{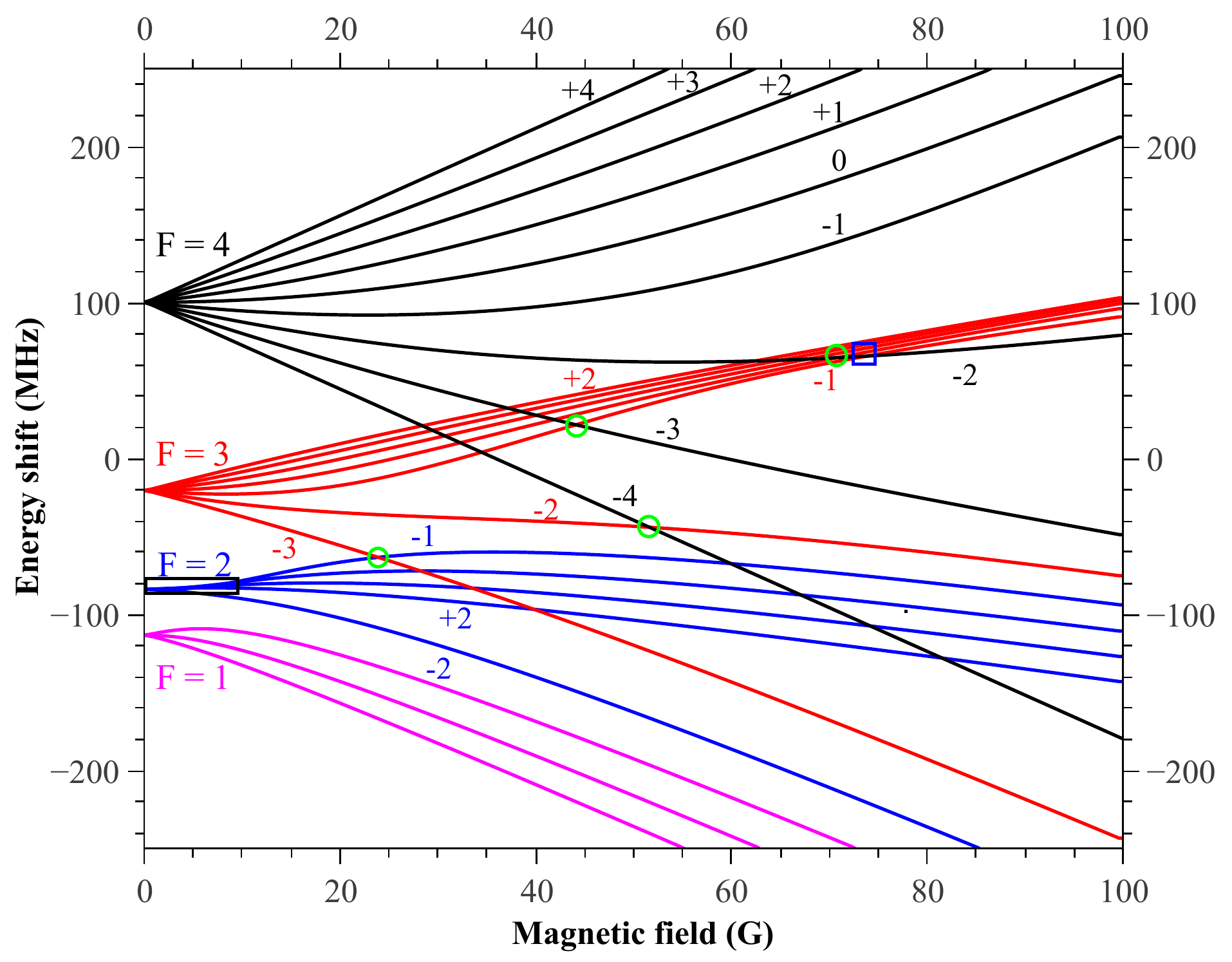}
    \caption{\label{fig:im3}(Color online) Energy shifts of the magnetic sublevels $m_F$ as a function of 
    magnetic field for $^{85}$Rb. Zero energy corresponds to the excited-state fine-structure level 5$^2$P$_{3/2}$. The numbers above the lines correspond to the values of $m_F$.
    Level crossings are marked by squares for $\Delta m_F=1$ and circles for $\Delta m_F=2$.
    }
    \end{figure}
\begin{figure}
	\includegraphics[width=\linewidth]{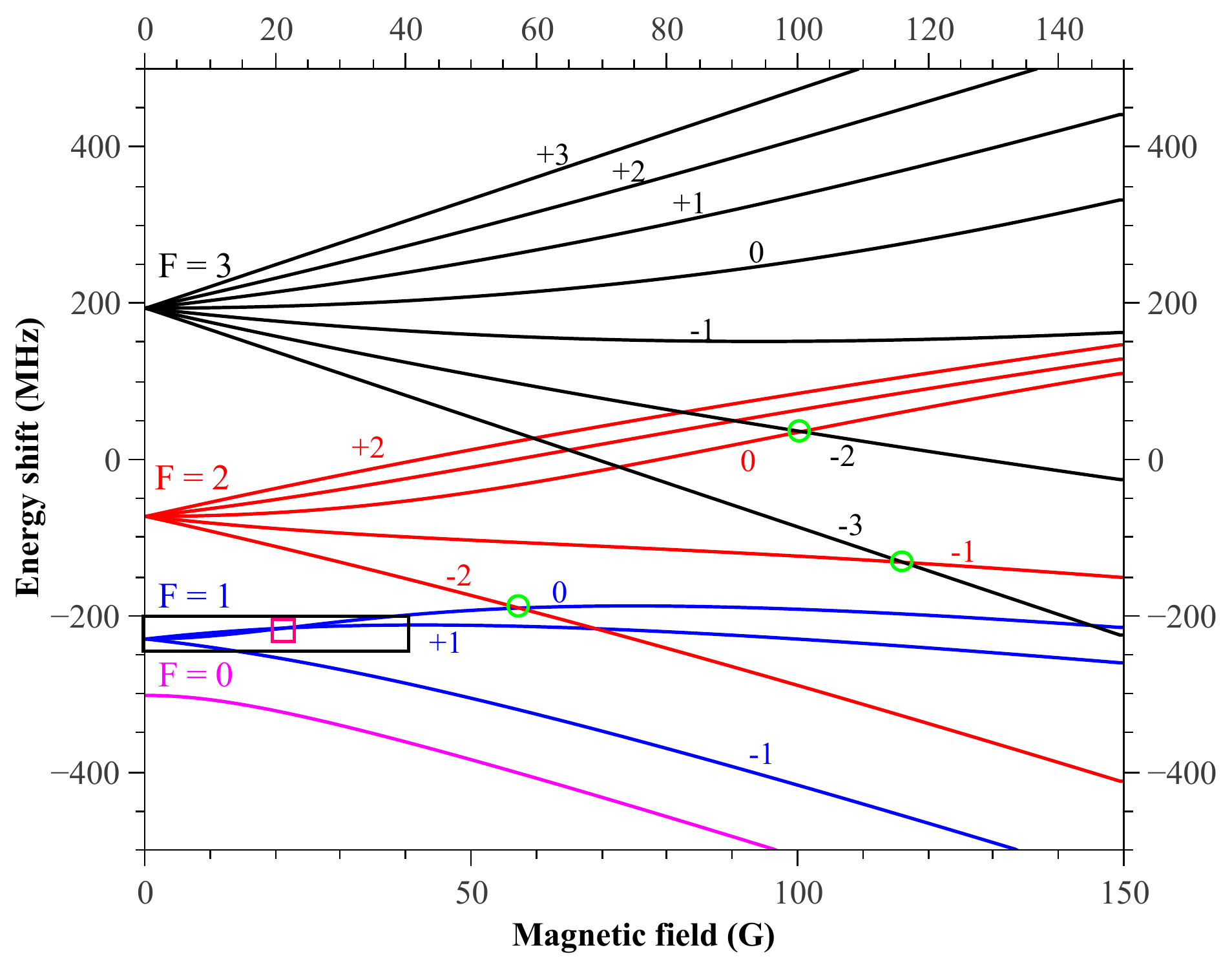}
    \caption{\label{fig:im5}(Color online) Energy shifts of the magnetic sublevels $m_F$ as a function of 
    magnetic field for $^{87}$Rb. Zero energy corresponds to the excited-state fine-structure level 5$^2$P$_{3/2}$. The numbers above the lines correspond to the values of $m_F$.
    Level crossings are marked by squares for $\Delta m_F=1$ and circles for $\Delta m_F=2$.
    }
\end{figure}
The frequency, direction, and polarization of light emitted from an ensemble of atoms is a sensitive probe of their 
quantum state~\cite{Auzinsh:2010book}. Changes in polarization such as, for example, rotation of the plane of polarization, 
are used to develop sensitive magnetometers~\cite{Budker:2002}. Other uses of nonlinear magneto-optical resonances include electromagnetically induced transparency~\cite{Harris:1997}, information storage using light~\cite{Phillips:2001, Liu:2001}, atomic clocks~\cite{Knappe:2005}, optical switches~\cite{Yeh:1982}, filters~\cite{Cere:2009}, and isolators~\cite{Weller:2012}.

When linearly polarized light interacts with an ensemble of atoms, it usually aligns the angular momentum of the atoms in the excited state as well as in the ground state. Angular momentum alignment can be symbolically represented by a double-headed arrow. If the angular momentum of the atoms is aligned along the quantization axis (longitudinal alignment), the populations of magnetic sublevels with quantum number $+m_F$ and $-m_F$ are equal, but the population may vary as a function of \(|{m_F}|\). But if the angular momentum is aligned perpendicularly to the quantization axis (transverse alignment), then, in quantum terms, it means that there is coherence between magnetic sublevels with quantum numbers that differ by $\Delta{m_F}=2$ (see Fig.~\ref{fig:im1}). 

In a similar way we can introduce longitudinal and transverse orientation of angular momentum. In the case of orientation of the angular momentum, the spatial distribution can be represented symbolically by a single-headed arrow, and in the case of longitudinal orientation, the magnetic sublevels with quantum numbers $+m_F$ and $-m_F$ in general have different populations. However, the case of transverse orientation corresponds to coherence between magnetic sublevels with values that differ by $\Delta{m_F}=1$ (see Fig.~\ref{fig:im1}).

The fluorescence from an aligned ensemble of atoms is expected to be linearly polarized, but in the case of oriented atoms, the fluorescence will possess a circularly polarized component as well.

Alignment created by linear polarized excitation can be converted to orientation by external interactions 
such as a magnetic field gradient~\cite{Fano:1964} or anisotropic collisions~\cite{Lombardi:1967,Rebane:1968,Manabe:1981}.
This process is called alignment-to-orientation conversion (AOC)\cite{Auzinsh:2005MolPol}. 
Interaction with an electric field also can produce orientation from an initially aligned 
population~\cite{Lombardi:1969}. 
A magnetic field by itself cannot create orientation from alignment because it is an axial field that is symmetric 
under reflection in the plane perpendicular to the field direction. 
However, the hyperfine interaction can cause a nonlinear dependence of the energies of the magnetic sublevels on the 
magnitude of the magnetic field---the nonlinear Zeeman effect (see Fig. \ref{fig:im3} and Fig. \ref{fig:im5}), 
and this nonlinear dependence can break the symmetry. If, in addition, the linearly polarized exciting radiation can be 
decomposed into linearly ($\pi^0$) and circularly ($\sigma^{\pm}$) polarized components 
with respect to the quantization axis (see Fig.~\ref{fig:im7}), then $\Delta m_F=1$ coherences can be created, which leads to 
orientation in a direction transverse to the initial alignment. 
AOC in an external magnetic field was first studied theoretically 
for cadmium~\cite{Lehmann:1964} and sodium~\cite{Baylis:1968}, and observed experimentally in cadmium~\cite{Lehmann:1969} and in the $D_2$ line of rubidium atoms~\cite{Krainska-Miszczak:1979}. Also the conversion in the opposite sense---conversion of an oriented state into an aligned---is possible~\cite{Weiss:1978}. 
Nevertheless, the action of external perturbations can break the symmetry of the population distribution and allow linearly 
polarized exciting radiation to produce orientation, which is manifested by the presence of circularly polarized fluorescence.
 
Earlier, AOC in rubidium atoms was studied at excitation with weak laser radiation in the linear absorption regime~\cite{Alnis:2001}. The perturbing factor in that case was the joint action of the hyperfine interaction and the external magnetic field, which led to nonlinear splitting of the Zeeman magnetic sublevels. The magnetic sublevels of the angular momentum hyperfine levels in Rb atoms in an external magnetic field start to be affected by the nonlinear Zeeman effect already at moderate field strengths of several tens of Gauss.

However, many practical and experimental applications require higher intensity excitation, in which case the absorption becomes nonlinear. 
As a result, the theoretical description is no longer simple and requires sophisticated methods in order to predict changes 
in the degree of circular polarization, which reaches maximum values on the order of only a few percent. 
Therefore, we have applied a theoretical model developed for the description of such magneto optical-effects like dark and bright resonances, to describe experimental signals of AOC 
in the $D_2$ line of rubidium. 
Because the splittings between the excited-state hyperfine levels are of the order of tens of 
megahertz for both rubidium isotopes (see Fig.~\ref{fig:im2}), the $D_2$ line is a very good candidate for 
demonstrating AOC phenomena at relatively low magnetic fields. 
The model satisfactorily calculates the degree of polarization for magnetic fields up to at least 85 Gauss, 
making it a powerful tool for experiments that deal with these effects.

We studied the AOC phenomenon experimentally by exciting the $D_2$ line of rubidium with linearly polarized light for the case of nonlinear absorption and modeled the line shapes of the resulting magneto-optical signals theoretically. 
Both circularly polarized components of the fluorescence were recorded in the experiment rather than just the difference as was done earlier~\citep{Alnis:2001}. 
Moreover, in the present study the magnetic field range was markedly extended in comparison 
to previous studies~\cite{Alnis:2001}, which allowed us to reveal additional signal structure.

\section{\label{sec:level2}Experiment}
\begin{figure}[h]
	\includegraphics[width=\linewidth]{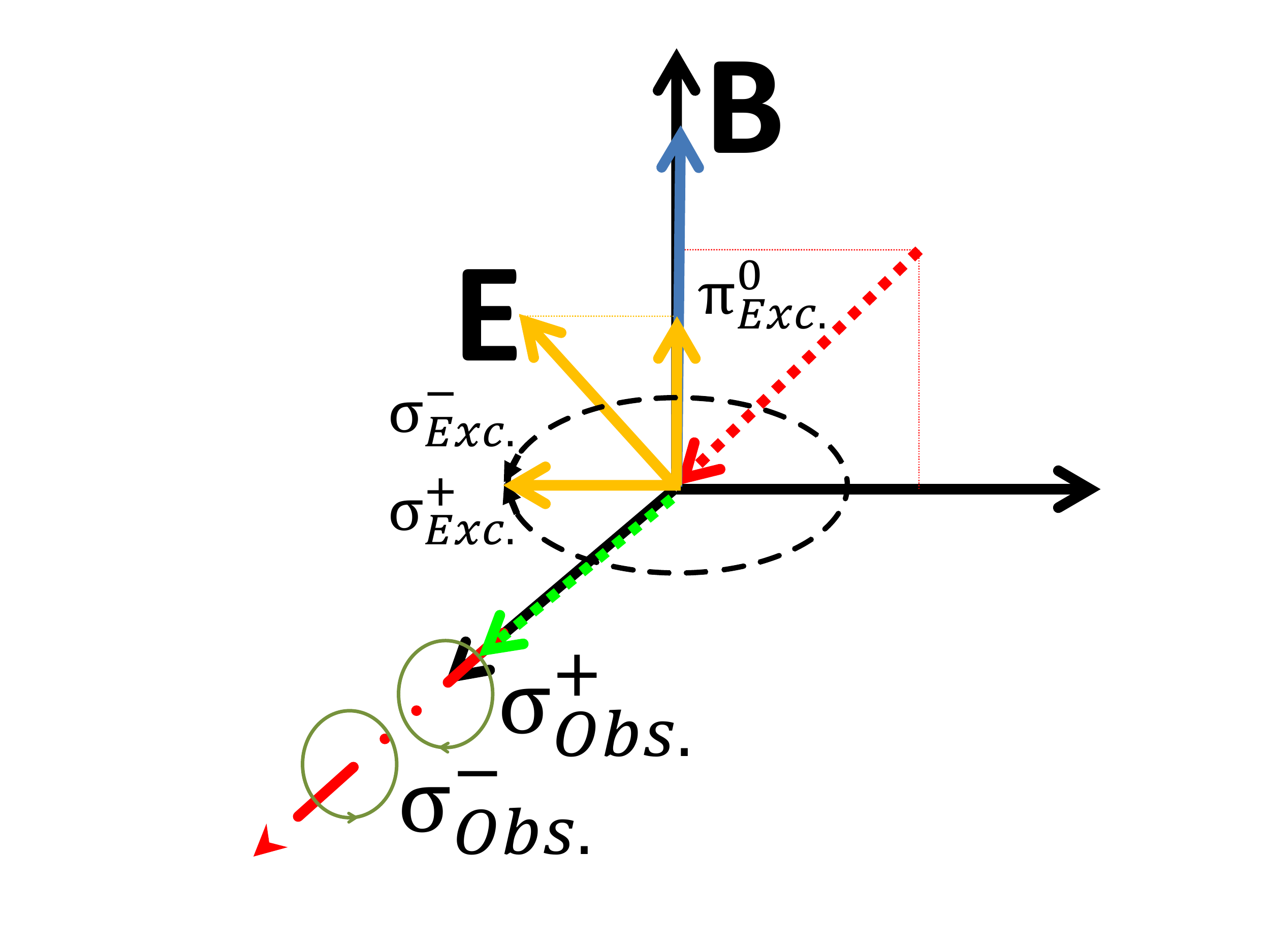}
    \caption{\label{fig:im7}(Color online) Excitation and observation geometry.}
\end{figure}

Rubidium atoms in a vapor cell were excited with linearly polarized light whose polarization vector made a 
45$^{\circ}$ angle with an externally applied magnetic field. Laser induced fluorescence (LIF) was observed 
in the direction perpendicular to the plane containing the magnetic field $\mathbf{B}$ and the electric field vector $\mathbf{E}$ of the 
exciting radiation (see Fig. \ref{fig:im7}) \citep{Auzinsh:1996}. The fluorescence in the observation direction passed through a 
two-lens system. Between the two lenses, a zero-order quarter-wave plate (Thorlabs WPQ10M-780) converted 
circularly polarized light into linearly polarized light. Next, a linear polarizer served as an analyzer, which 
allowed one or another circularly polarized fluorescence component to pass, depending on the relative angle between the 
analyzer axis and the fast axis of the quarter-wave plate. 
 
\begin{figure}[h]
	\includegraphics[width=\linewidth]{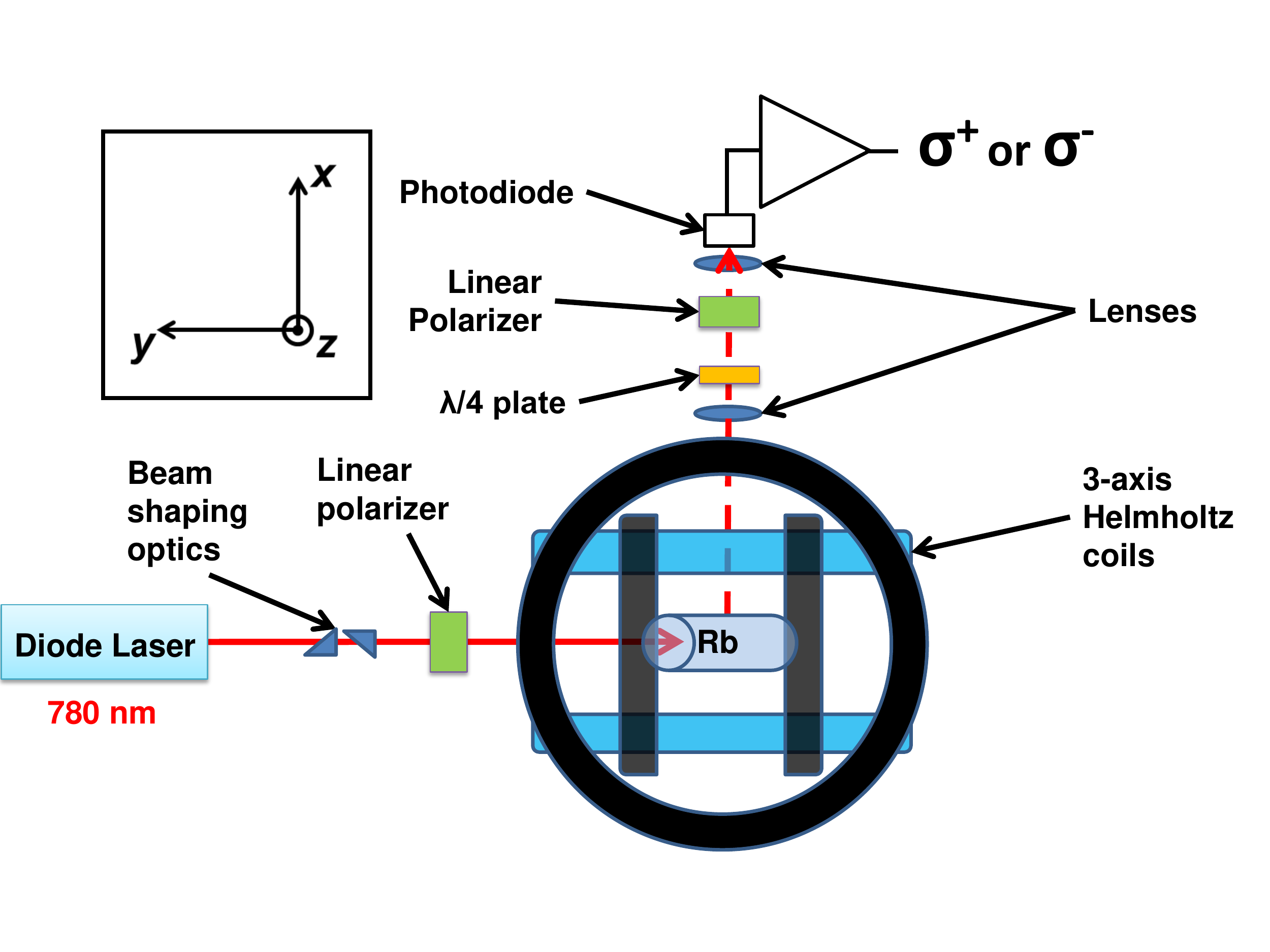}
    \caption{\label{fig:im8}(Color online) Top view of the experimental setup. Although in the top view it appears that the beam is parallel to the $y$ axis, in fact it enters the coils at an angle of 45$^{\circ}$ with respect to $y$ axis in the $yz$ plane}
\end{figure}
\indent The experimental apparatus is shown schematically in Fig. \ref{fig:im8}. 
Rubidium atoms from a natural isotopic mixture were contained in a cylindrical Pyrex cell (length and diameter both 25 mm) 
with optical quality windows.  The rubidium cell was located at the center of three pairs of mutually orthogonal 
Helmholtz coils. The magnetic field was scanned in the $z$-direction, while the two remaining coils 
were used to compensate the ambient static magnetic field. 
We estimate that the ambient magnetic field was compensated to better than 0$\pm$20 mG. 
In order to scan the magnetic field in both directions, a bipolar power supply (Kepco BOP-50-8-M) was used, 
reaching magnetic field values of 85 G in both directions.

The laser used in these experiments was a Toptica DL Pro grating-stabilized, tuneable, single-mode diode laser. 
The frequency of the laser excitation was stabilized by generating a saturated absorption spectrum and locking the laser frequency to a saturated absorption peak in this signal using a Toptica DigiLock 110 feedback control module. 
The frequency was additionally monitored by a HighFinesse WS/7 Wavemeter. 
The temperature and current of the laser were controlled by Toptica DTC 110 and DCC 110 controllers, respectively.

The diameter of the beam was 1.90~mm at the full width at half maximum (FWHM) as determined from the Gaussian fit obtained 
by a beam profiler (Thorlabs BP104-VIS). The ellipticity of the laser beam was compensated by an anamorphic prism pair. 
The laser power was changed using neutral density filters placed before the linear polarizer. 
The LIF of the two opposite circularly polarized light components was collected on a photodiode (Thorlabs FDS100). 
Each component was measured separately and multiple scans were acquired and averaged before switching the analyzing  
polarizer in order to measure the orthogonally polarized component. The signal was amplified by a transimpedance 
amplifier based on a TL072 op-amp with a gain of 10$^6$ followed by a voltage amplifier 
with a gain of 10$^4$ (Roithner multiboard). 
The signals were stored after each scan on a PC using an Agilent DSO5014A oscilloscope. 
A residual misalignment in the experimental setup introduced a slight asymmetry in the signal, but it could be eliminated by 
averaging the signals recorded for positive and negative values of magnetic field.

In order to compare experiment with theory, both components were normalized to the maximum of the $\sigma^+$ component, making it possible to compare the relative intensities of the two components in arbitrary units. The background was measured in two different ways: by detuning the laser frequency from resonance and by blocking the laser beam. Both produced equal results. In the fitting process a constant background was introduced, which was close to the experimentally measured background. 
The experimental results were very sensitive to any slight misalignment of the analyzing polarizer that could distort the measured strengths of each circular polarization component. Therefore, to find the best agreement between experiment and theory, a parameter was varied that represented the 
relative strength of each experimentally measured fluorescence component. This factor was usually around 10$\%$ and never more than 22$\%$. 

\section{\label{sec:level3}Theoretical model}
A well-tested model based on optical Bloch equations (OBEs) that are solved for steady state excitation conditions is used to describe 
the experiment theoretically. The ensemble of rubidium atoms is described by a quantum density matrix $\rho$ 
that is written in the basis $\xi, F_i, m_{F_i}$, where $F_i$ denotes the quantum number of the total atomic 
angular momentum including nuclear spin $I$ for either the ground ($i = g$) or the excited ($i = e$) state, 
$m_{F_i}$ is the magnetic quantum number and $\xi$ stands for all the other quantum numbers that are irrelevant 
in the context of our experiment. Thus, the general OBEs \cite{Stenholm:2005},
\begin{equation}\label{eq:liouville}
	i\hbar\frac{\partial \rho}{\partial t} = \left[\hat H,\rho\right] + i\hbar\hat R\rho,
\end{equation}
can be transformed into explicit rate equations for the Zeeman coherences within the ground ($\rho_{g_ig_j}$) 
and excited ($\rho_{e_ie_j}$) states, respectively. 
To do so, the laser radiation is described as a classically oscillating electric field 
$\mathbf{E}(t)$ with a stochastically fluctuating phase. Thus, the interaction operator can be written in the dipole approximation with dipole operator $\hat{\mathbf{d}}$.
\begin{equation}\label{eq:v-operator}
	\hat V = -\hat{\mathbf{d}}\cdot \mathbb{E}(t)
\end{equation}
The interaction with the magnetic field is described by the operator
\begin{equation}\label{eq:zeeman-effect}
	\hat H_B = \frac{\mu_B}{\hbar}\left(g_J\mathbf{J} + g_I\mathbf{I}\right)\cdot \mathbf{B},
\end{equation}
where $\mathbf{J}$ and $\mathbf{I}$ are, respectively, the total electronic angular momentum and nuclear spin, which 
together make up the total atomic angular momentum $\mathbf{F}$. 
The quantities $g_J$ and $g_I$ are the respective Land\'{e} factors, 
$\mathbf{B}$ is the external magnetic field, $\mu_B$ is Bohr's magneton, and $\hbar$ is Planck's constant. 
The matrix elements for the electric dipole transition can be written in explicit matrix form with the help of Wigner-Eckart 
theorem~\cite{Auzinsh:2009b}.

Thus the total interaction Hamiltonian in (\ref{eq:liouville}) is
\begin{equation}
	\hat H = \hat H_0 + \hat H_B + \hat V,
\end{equation}
where $H_0$ governs the internal energy of an unperturbed atom. 

The relaxation operator in (\ref{eq:liouville}) includes terms for the spontaneous relaxation rate $\Gamma$ 
and the transit relaxation rate $\gamma$, which is the inverse of the average time an atom takes to traverse the laser beam.

By applying the rotating wave approximation, averaging over and decorrelating from the stochastic phases, 
and eliminating the optical coherences as described in detail by Blush and Auzinsh~\cite{Blushs:2004}, 
the rate equations for the Zeeman coherences are obtained:
\begin{subequations} \label{eq:zc}
\begin{align}
\frac{\partial \rho_{g_ig_j}}{\partial t} =& \left(\Xi_{g_ie_m} + \Xi_{g_je_k}^{\ast}\right)\sum_{e_k,e_m}d_{g_ie_k}^\ast d_{e_mg_j}\rho_{e_ke_m} - \nonumber\\ -& \sum_{e_k,g_m}\Big(\Xi_{g_je_k}^{\ast}d_{g_ie_k}^\ast d_{e_kg_m}\rho_{g_mg_j} + \nonumber\\ +& \Xi_{g_ie_k}d_{g_me_k}^\ast d_{e_kg_j}\rho_{g_ig_m}\Big) -  i \omega_{g_ig_j}\rho_{g_ig_j} -\nonumber\\-& \gamma\rho_{g_ig_j} + \sum_{e_ke_l}\Gamma_{g_ig_j}^{e_ke_l}\rho_{e_ke_l} + \lambda\delta(g_i,g_j) \label{eq:zcgg} \\
\frac{\partial \rho_{e_ie_j}}{\partial t} =& \left(\Xi_{g_me_i}^\ast + \Xi_{g_ke_j}\right)\sum_{g_k,g_m}d_{e_ig_k} d_{g_me_j}^\ast\rho_{g_kg_m} -\nonumber\\-& \sum_{g_k,e_m}\Big(\Xi_{g_ke_j}d_{e_ig_k} d_{g_ke_m}^\ast\rho_{e_me_j} + \nonumber\\ +& \Xi_{g_ke_i}^\ast d_{e_mg_k} d_{g_ke_j}^\ast\rho_{e_ie_m}\Big) -  i \omega_{e_ie_j}\rho_{e_ie_j} -\nonumber\\ - & (\Gamma + \gamma)\rho_{e_ie_j}. \label{eq:zcee}
\end{align}
\end{subequations}
The first two terms in both equations describe the population increase/decrease and the creation of 
Zeeman coherences within the respective atomic states due to the interaction of atoms with the laser radiation. 
The elements of the transition dipole matrix are given by $d_{ij}$ [obtained from ({\ref{eq:v-operator})], 
and $\Xi_{ij}$, which is defined below in equation (\ref{eq:xi}), gives the atom-field interaction strength. 
The third term of the rate equations (\ref{eq:zc}) describes the destruction of coherence by the magnetic field, and 
$\omega_{ij}$ is the energy difference between magnetic sub-levels $\vert i\rangle$ and $\vert j\rangle$ and can be obtained by 
diagonalizing the matrix $\hat H_0 + \hat H_B$. 
The fourth term describes the population loss and destruction of coherence caused by relaxation. 
The fifth term in (\ref{eq:zcgg}) describes repopulation of the ground state by spontaneous transitions and the sixth, 
repopulation by transit relaxation. 
If we assume that the atomic density matrix outside the interaction region is normalized, 
then $\lambda = \frac{1}{n_g}\gamma$, where $n_g$ is the total number of magnetic sub-levels in the ground state.

The quantity $\Xi_{g_ie_j}$ in equation \eqref{eq:zc} describes the strength of interaction between the laser radiation and the atoms and is expressed as follows:
\begin{equation}\label{eq:xi}
\Xi_{g_ie_j} = \frac{\Omega_{R}^{2}}{\frac{\Gamma+\gamma+\Delta\omega}{2}+\dot\imath\left(\bar\omega-\mathbf{k}_{\bar\omega}\cdot\mathbf{v}
	+\omega_{g_ie_j}\right)},
\end{equation}
where $\Omega_{R}$ is the reduced Rabi frequency, used as a theoretical parameter that corresponds to the 
laser power density in the experiment, 
$\Delta\omega$ is the finite spectral width of the exciting radiation, 
$\bar \omega$ is the central frequency of the exciting radiation, 
$\mathbf{k}_{\bar\omega}$ the wave vector of exciting radiation, 
and $\mathbf{k}_{\bar\omega}\mathbf{v}$ is the Doppler shift experienced by an atom moving with velocity $\mathbf{v}$.

During the experiment steady interaction conditions were maintained. 
Thus, we can apply steady state conditions
\begin{equation}\label{eq:steady}
	0 = \frac{\partial \rho_{g_ig_j}}{\partial t} = \frac{\partial \rho_{e_ie_j}}{\partial t},
\end{equation}
obtaining from (\ref{eq:zc}) a set of linear equations 
that can be solved numerically to obtain the density matrix components that correspond to the population and the Zeeman coherences 
of the ground and excited states. Once the density matrix is known we use the following expression to obtain 
the intensity (up to a constant factor $\tilde{I}_0$) of an arbitrary polarized fluorescence component with polarization 
denoted by $\mathbf{e}$:
\begin{equation}\label{eq:fluorescence}
	I_{fl}(\mathbf{e}) = \tilde{I}_0\sum\limits_{g_i,e_j,e_k} d_{g_ie_j}^{\ast(ob)}d_{e_kg_i}^{(ob)}\rho_{e_je_k}.
\end{equation}
To include the effects of the thermal motion of the atoms, we perform Riemann integration over the velocity distribution by solving Eqs. (\ref{eq:zc}) and evaluating (\ref{eq:fluorescence}) for each atomic velocity group.

To fit the theoretical and experimental results we estimate and fine tune the following parameters: transit relaxation rate $\gamma$, reduced Rabi frequency $\Omega_R$, and spectral width of the laser radiation $\Delta\omega$.

The estimation of the transit relaxation rate is straightforward:
\begin{equation}\label{eq:gamma}
	\gamma = \frac{v_{th}}{d},
\end{equation}
where $v_{th}$ is the mean thermal velocity of the atoms projected on to the plane perpendicular to the laser beam and $d$ is the laser beam diameter, which in the theoretical model is assumed to be cylindrical in shape with uniform power density. 
For $d = 1900~\mu$m 
and $T = 293$~K we obtain $\gamma = 2 \pi \cdot (0.018$~MHz$)$.

The Rabi frequency can be estimated theoretically as
\begin{equation}\label{eq:Rabi}
	\Omega_R = k_{R}\frac{\vert\vert d\vert\vert\cdot\vert\varepsilon\vert}{\hbar} = k_{R}\frac{\vert\vert d\vert\vert}{\hbar}\sqrt{\frac{2I}{c}},
\end{equation}
where $k_R$ is some fitting parameter of order unity, 
$\vert\vert d\vert\vert$ is the reduced dipole matrix element that remains unchanged 
for all transitions within the $D_2$ line~\cite{Auzinsh:2009b}, 
$I$ is the laser power density (directly related to the amplitude of the electric field $\vert\varepsilon\vert$), 
and $c$ is the speed of light. 
In practice, the estimation is not straightforward as the power density $I$ is not constant across the laser beam, but in 
the theoretical model only a constant average value is used in place of the actual power distribution. 
Theoretical and experimental evidence suggests \cite{Fescenko:2012, Auzinsh:2015} that $\Omega_R$ cannot be related to the square root of the laser power density $I$ by a simple constant $k_R$ for all values of the laser power density if one merely 
assumes that the power density distribution within the beam is Gaussian.

This fact has a simple explanation. Our experiment was performed in the 
regime of nonlinear absorption, which implies that for large laser intensities the ground state population is strongly depleted. When one starts to gradually increase laser intensity, initially the ground-state population is only slightly changed even at the center of the beam, where the light is most intense. When the intensity is increased still more, the ground-state population at the center of the beam starts to be  depleted significantly. When the intensity is increased further, there is little ground-state population left in the beam center, and the region of population depletion expands to the ``wings'' of the Gaussian laser power density distribution, which can extend a significant distance from the laser beam's center. 

As a consequence, although the theoretical proportionality of  $\Omega_R$ to the square root of laser power density holds, for weaker laser radiation the main contribution to the signal comes from the central parts of the laser beam where we have the strongest power density. In contrast, for stronger laser radiation the role of the peripheral parts of the laser beam, where the radiation power density is smaller, starts to play a larger role in the absorption process, because only there the ground-state population is still significant. In each of these cases the radiation power density in different parts of the beam plays a dominant role in the absorption process and should be related to value of the Rabi frequency that appears in the rate equations for the density matrix. Thus we vary the value of coefficient $k_R$ in order to account for this effect and to achieve better correspondence between experiment and theory. 

A value of $\Delta\omega = 2\pi\cdot(1$~MHz$)$ was found to be an appropriate estimate for the spectral width of the laser and is close to the value given by the manufacturer of the laser.

\section{\label{sec:level4}Results and discussion}
Before the experiments were carried out, some preliminary theoretical calculations were performed in order to deduce which 
hyperfine transition would yield the most noticeable signals related to the AOC phenomenon in both rubidium isotopes.
A good measure of the strength of the AOC effect is the degree of circularity of the laser induced fluorescence, defined as 
($I_{\sigma^+}$ -- $I_{\sigma^-}$)/($I_{\sigma^+}$ + $I_{\sigma^-}$).
The theoretical calculations predicted that the largest circularity signal (4$\%$) would be observed for $^{85}$Rb when excited from the second ground-state hyperfine level $F_g=2$ to the second excited-state hyperfine level $F_e=2$.  
\begin{figure*}[p]
	\includegraphics[width=0.3297\textwidth]{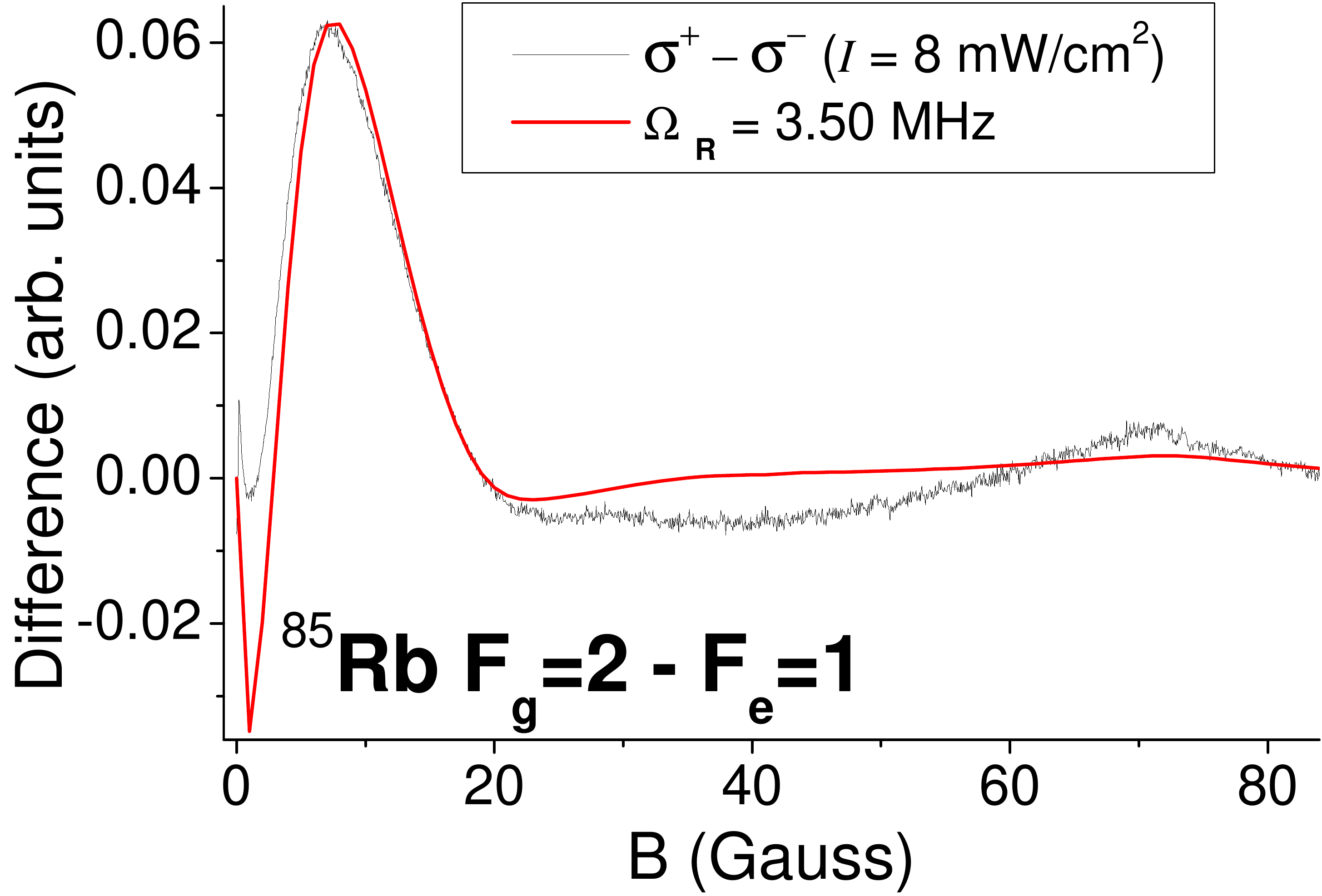}
	\includegraphics[width=0.3297\textwidth]{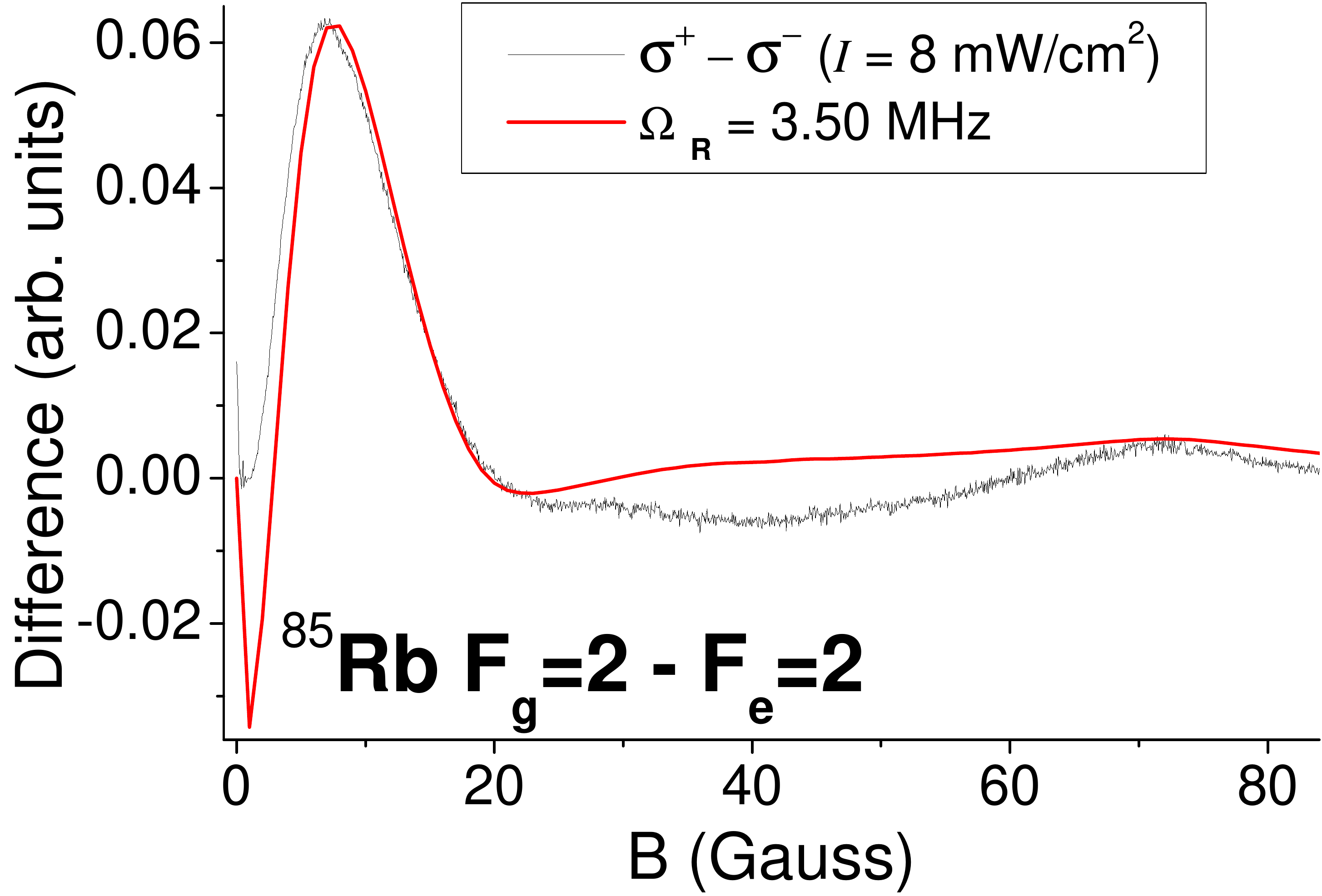}
	\includegraphics[width=0.3297\textwidth]{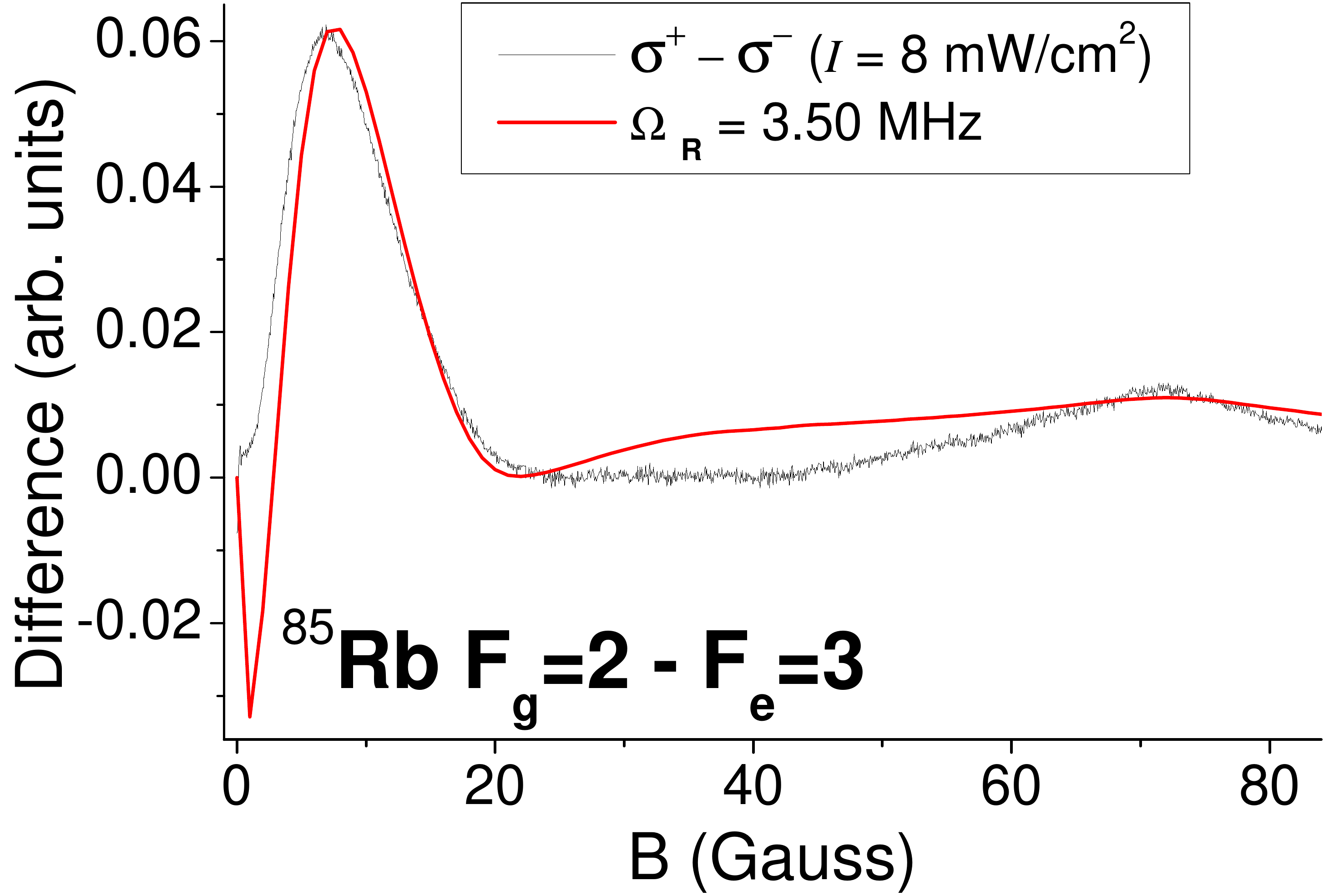}
   \caption{\label{fig:im10}(Color online) Signal dependence on the excited-state hyperfine level $F_e$ to which the laser is tuned when excited from the ground-state hyperfine level $F_g=2$ of $^{85}$Rb. 
   The left-most plot shows the difference between the two oppositely circularly polarized components ($I_{\sigma^+}-I_{\sigma^-}$) for the $F_g=2 \longrightarrow F_e=1$ transition, the center plot shows the difference ($I_{\sigma^+}-I_{\sigma^-}$) for the $F_g=2 \longrightarrow F_e=2$ transition, and the third plot (right-most) corresponds to the $F_g=2 \longrightarrow F_e=3$ transition.}
\end{figure*}
As seen in Fig.~\ref{fig:im10}, because of Doppler broadening, the signal did not depend significantly on which excited-state hyperfine level was excited when the excitation took place from the ground-state hyperfine level with $F_g=2$. 
The observable circularity for the other transitions was 
predicted to be 1\% or less. 
For the case of $^{87}$Rb, the $F_g=1\longrightarrow F_e=1$ transition was selected, because the predicted circularity degree was 1$\%$, whereas for excitation from the other ground-state hyperfine level $F_g=2$, the circularity degree was predicted to be less than 1$\%$. Therefore we concentrated our experimental efforts on the $F_g=2\longrightarrow F_e=2$ transition of $^{85}$Rb and the $F_g=1\longrightarrow F_e=1$ transition of $^{87}$Rb.\\
\begin{figure*}[t]
	\includegraphics[page=1,width=\linewidth]{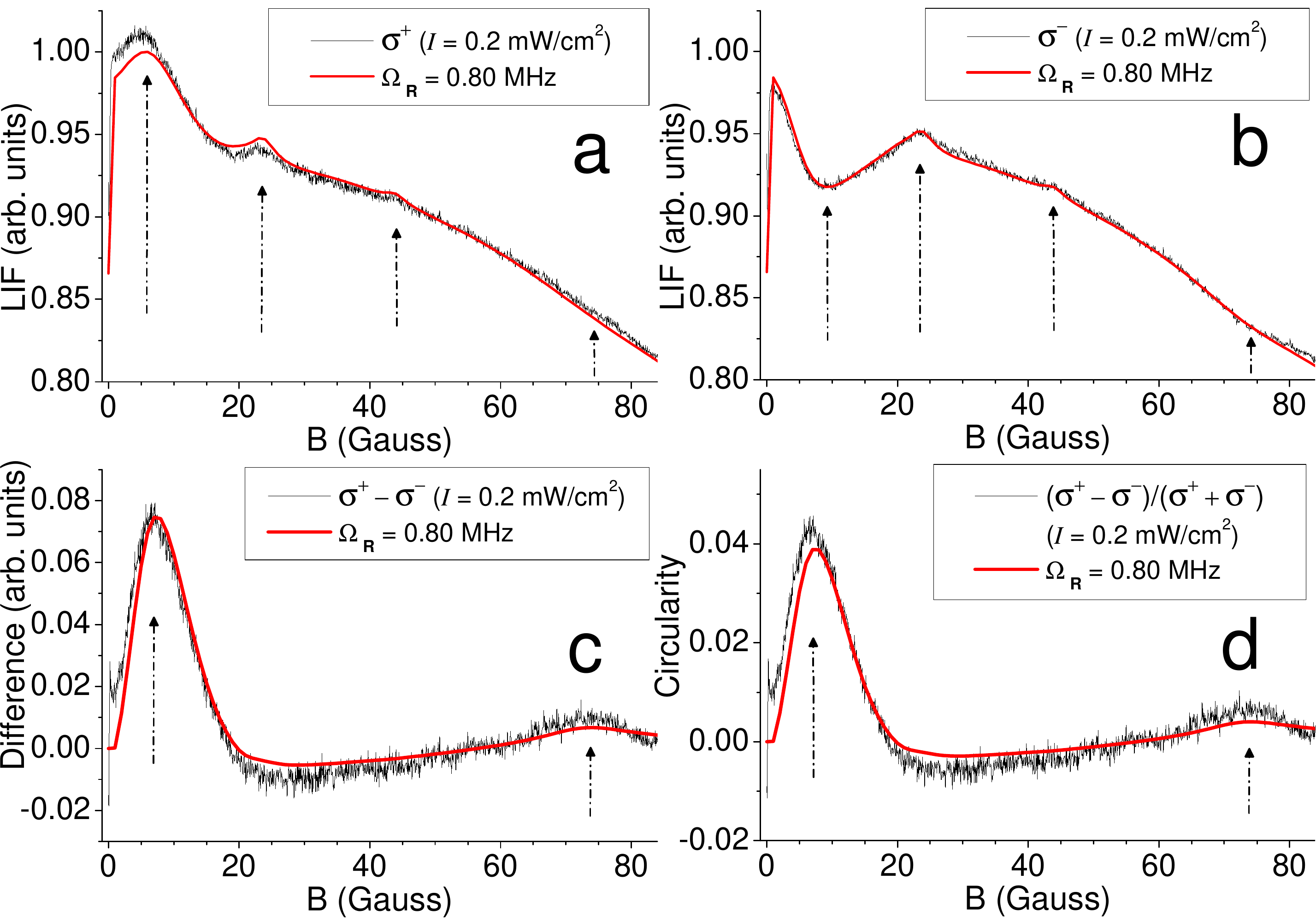}
	\caption{\label{fig:im9}(Color online) Relative intensities of the two oppositely circularly polarized fluorescence componentes 
	(a) and (b), their difference (c), and the circularity value (d) for the $F_g=2 \longrightarrow F_e=2$ transition of $^{85}$Rb
	(80 scans averaged).  
	Arrows denote the positions of peaks and maximum (or minimum) values of broader structures.}
\end{figure*}
\begin{figure}[h]
	\includegraphics[width=\linewidth]{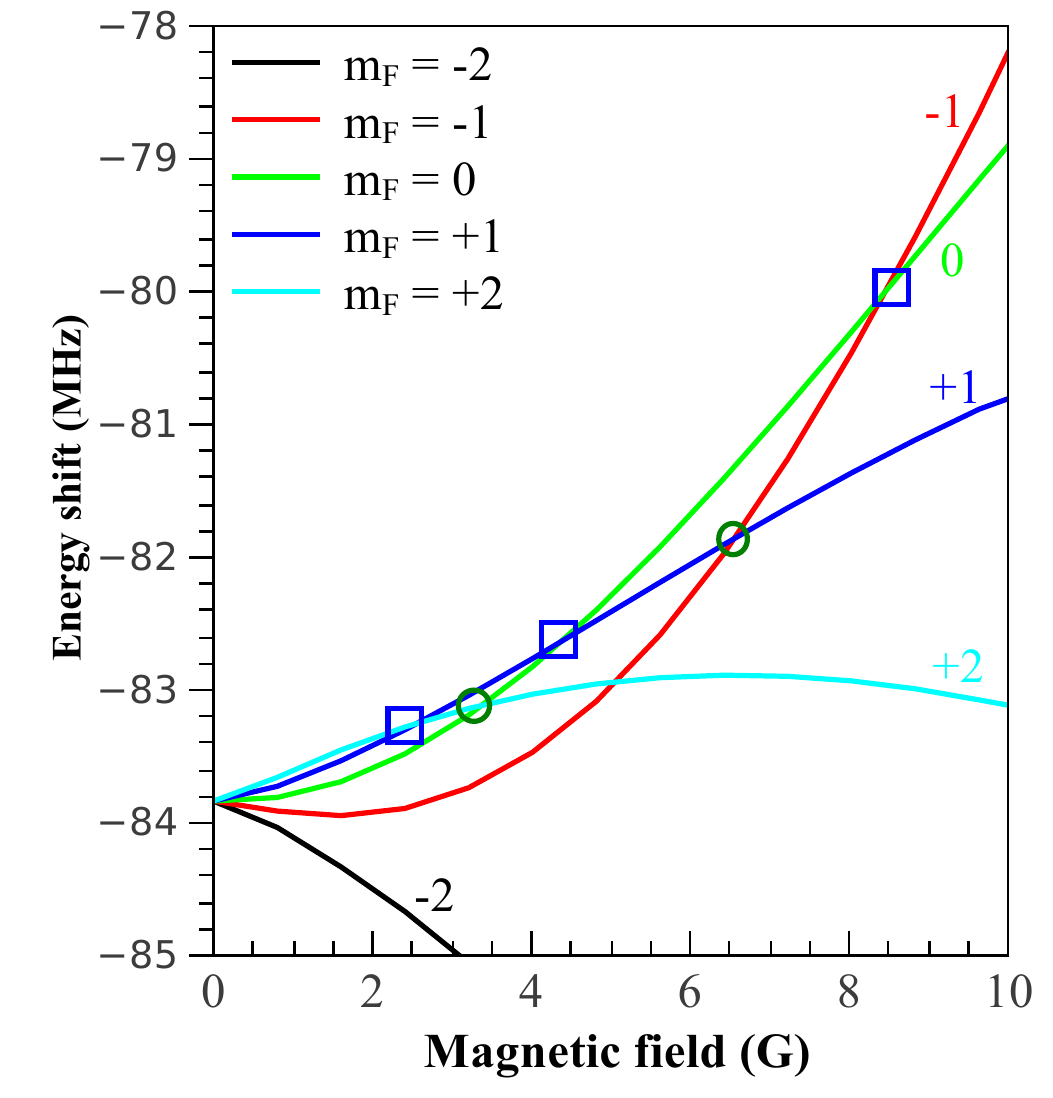}
    \caption{\label{fig:im4}(Color online) Energy shifts of the magnetic sublevels $m_F$ as a function of magnetic 
    field for $^{85}$Rb in the magnetic field range $0\ \text{G} < B < 10\ \text{G}$. The $m_F$ values are written next to the curves. Squares denote $\Delta m_F=1$ crossings and circles denote $\Delta m_F=2$ crossings.}
\end{figure}
\indent Figure \ref{fig:im9} shows a typical result for the $F_g=2 \longrightarrow F_e=2$ transition of $^{85}$Rb. 
Figure~\ref{fig:im9}(a) and Fig. \ref{fig:im9}(b) depict the two orthogonally circularly polarized fluorescence components. 
When the magnetic field value is zero, all magnetic sublevels $m_F$ that belong to the same $F$ level in the excited and ground states are degenerate, 
giving a typical dark resonance for the 
$F_g=2 \longrightarrow F_e=2$ transition of $^{85}$Rb~\cite{Auzinsh:2009}. 
As the magnetic field magnitude increases, these sublevels shift according to the nonlinear Zeeman effect 
(Fig.~\ref{fig:im3}), thereby destroying the aligned state and allowing more laser light to be absorbed, 
which causes a rapid rise in the fluorescence signal. 
After that the overall signal tendency is to diminish as the magnetic field strength increases, 
apart from two small peaks at about 23 G and 44 G.

These two small peaks can be attributed to $\Delta m_F=2$ coherences. 
The 23 G peak appears because the $m_F=-1$ sublevel of the $F_e=2$ 
hyperfine level crosses the $m_F=-3$ sublevel of the \textit{F$_e$}=3 hyperfine level 
(see Fig.~\ref{fig:im3}), thus creating a $\Delta m_F=2$ coherence. 
The other small peak at 44 G can be attributed to the crossing of $m_F=-1$ sublevel of the $F_e=3$ and the 
$m_F=-3$ sublevel of $F_e=4$. 
Note that these peaks are invisible both in the difference signal [Fig.~\ref{fig:im9}(c)] 
as well as in the circularity signal [Fig.~\ref{fig:im9}(d)] since they cancel each other when the difference is taken.

Besides these two small peaks in the component graphs, there are two peaks at 7 G and 74 G in the difference and circularity graphs [Fig.~\ref{fig:im9}(c) and (d)] corresponding to the two broader structures in the component graphs [Fig.~\ref{fig:im9}(a) and (b)]: one around 6--10~G, and another, barely visible one around 70--74~G.
These peaks can be 
\begin{figure*}
	\includegraphics[page=1,width=\textwidth]{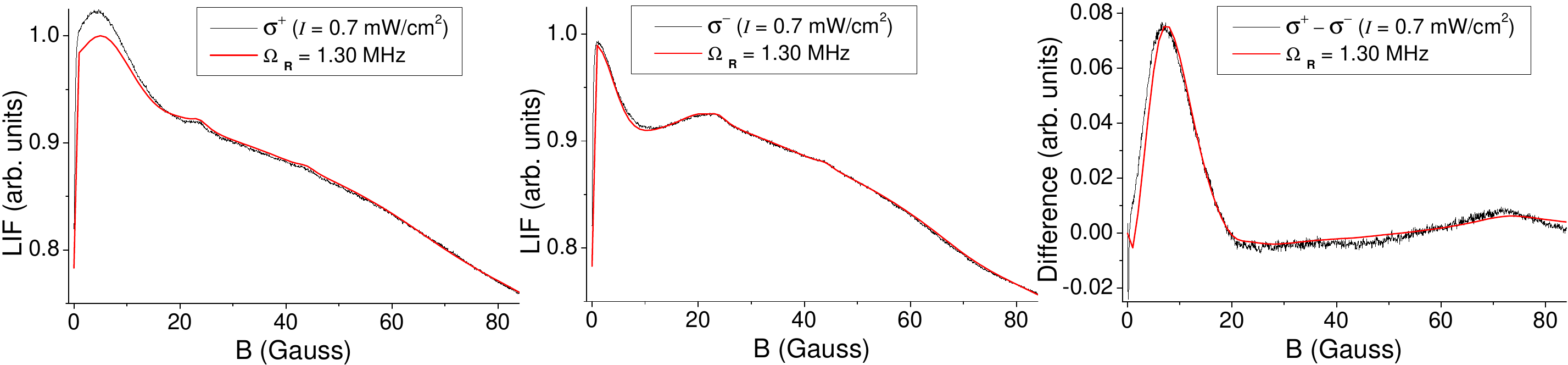}
	\includegraphics[page=2,width=\textwidth]{Rb85_Fg2_Fe2_Power_1x3}
	\includegraphics[page=3,width=\textwidth]{Rb85_Fg2_Fe2_Power_1x3}
	\caption{\label{fig:im11}(Color online) Signal dependence on laser power density for excitation of the 
	$F_g=2 \longrightarrow F_e=2$ transition of the $D_2$ line of $^{85}$Rb. 
	The plots are organized in columns: relative intensities of the two oppositely circularly polarized fluorescence components are shown in the left and center columns and their difference in the right-most column.}
\end{figure*}
\begin{figure*}
	\includegraphics[page=1,width=\textwidth]{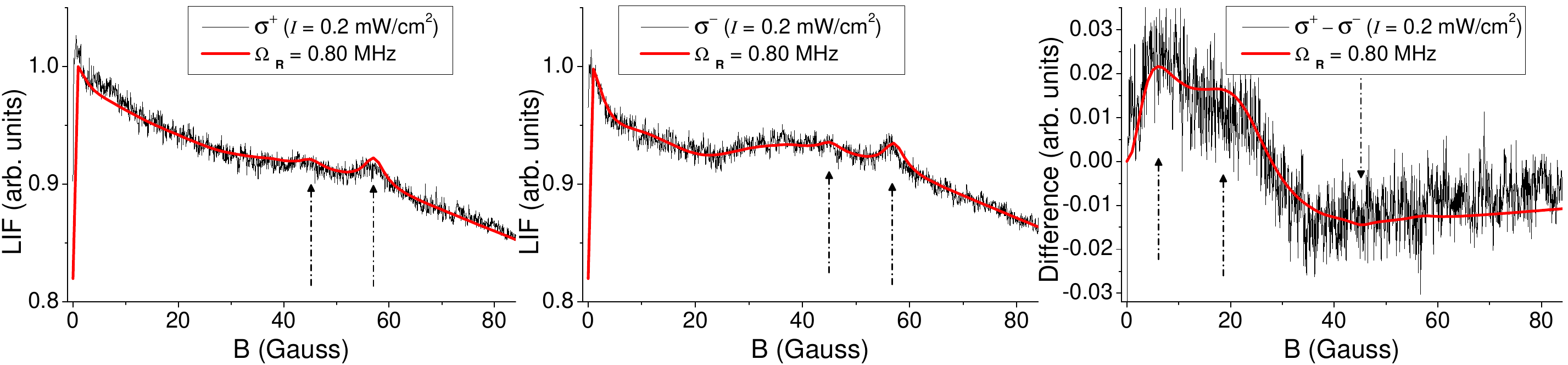}
	\includegraphics[page=2,width=\textwidth]{Rb87_Fg1_Fe1_Power_1x3}
	\includegraphics[page=3,width=\textwidth]{Rb87_Fg1_Fe1_Power_1x3}
	\includegraphics[page=4,width=\textwidth]{Rb87_Fg1_Fe1_Power_1x3}
	\caption{\label{fig:im12}(Color online) Signal dependence on laser power density for excitation of the 
	$F_g=1 \longrightarrow F_e=1$ transition of the $D_2$ line of $^{87}$Rb. 
	The plots are organized in columns: relative intensities of the two oppositely circularly polarized fluorescence components are shown in the left and center columns and their difference in the right-most column. Arrows denote the positions of peaks and maximum values of broader structures.}
\end{figure*}
\begin{figure}[h]
	\includegraphics[width=\linewidth]{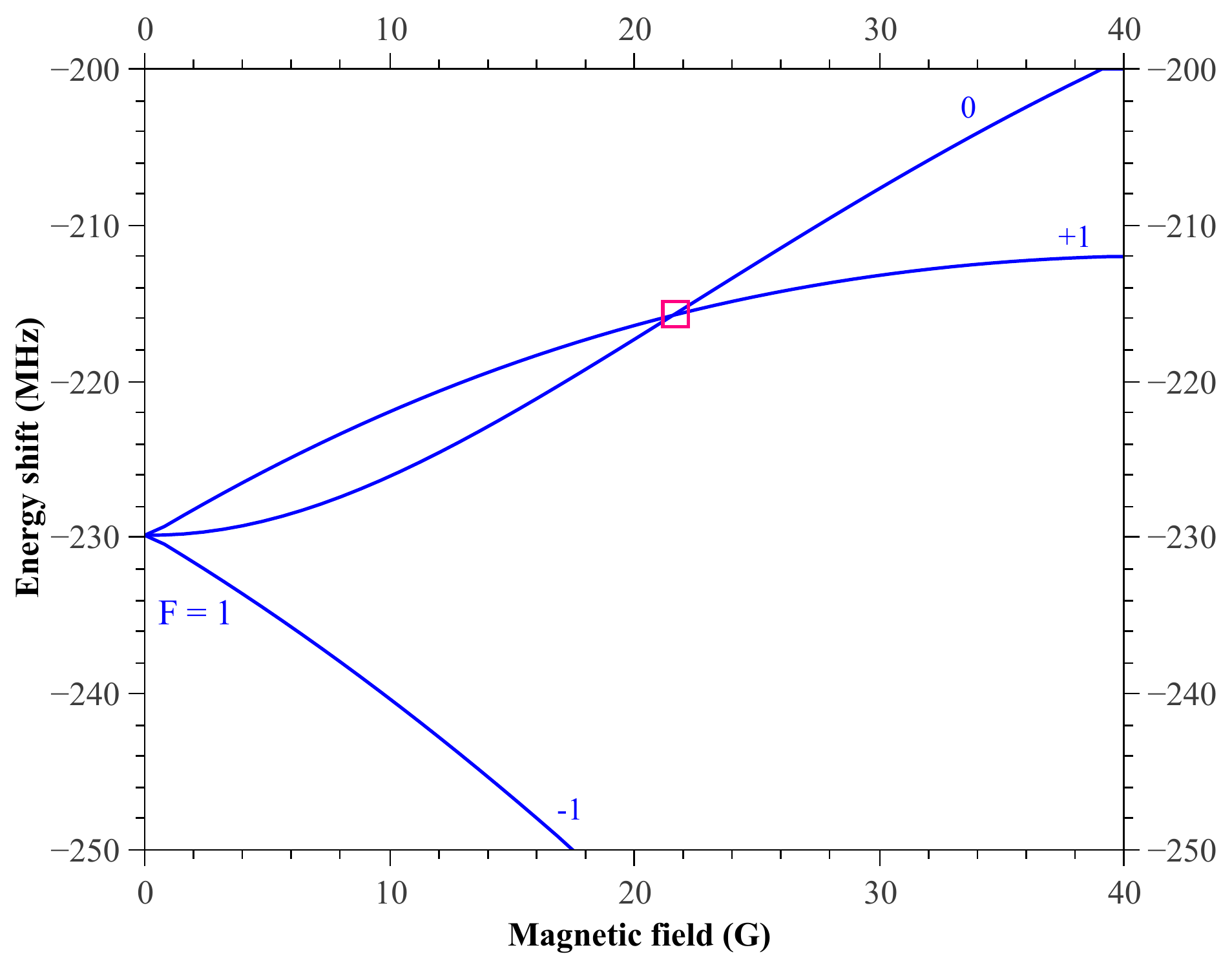}
    \caption{\label{fig:im6}(Color online) Energy shifts of the magnetic sublevels $m_F$ as a function of magnetic field for $^{87}$Rb 
    in the magnetic field range $0\ \text{G} < B < 40\ \text{G}$. The $m_F$ values are written next to the curves. The square denotes a $\Delta m_F=1$ crossing.}
\end{figure}

attributed to $\Delta m_F=1$ coherences. The 7 G peak appears as an increase in the signal in one component 
[Fig.~\ref{fig:im9}(a)] and a decrease in the other [Fig.~\ref{fig:im9}(b)]. Note that their corresponding 
maximum and minimum values are relatively shifted, giving values of 6 G [Fig.~\ref{fig:im9}(a)] and 9 G 
[Fig.~\ref{fig:im9}(b)], respectively, in the component graphs. The relative shift of these values can be 
explained by the fact that this peak is related to three $\Delta m_F=1$ and two $\Delta m_F=2$ coherences in the range 
from 0 to 10 G (see Fig.~\ref{fig:im4}). As we take the difference between the two oppositely circularly 
polarized components, we can eliminate the $\Delta m_F=2$ coherences from the signal and thus see the peaks that 
correspond only to the $\Delta m_F=1$ crossings. The 74 G peak in Fig.~\ref{fig:im9}(c) can be explained 
in a similar way. A barely visible structure in the component graphs appears as a broad peak 
in the difference graph. This peak is related to a single $\Delta m_F=1$ crossing of the $m_F=-1$ 
sublevel of $F_e=3$ and the $m_F=-2$ sublevel of $F_e=4$, and as a result its amplitude is smaller. 
The peak is broad because the $m_F=-1$ and $m_F=-2$ sublevels that cross are energetically close 
to each other ($\Delta{E} \le 20$ MHz ) all the way from 60 G to 90 G, as can be seen in Fig.~\ref{fig:im3}.

Figure~\ref{fig:im11} shows the signal dependence on laser power for the $F_g=2 \longrightarrow F_e=2$ transition. 
One can see in the figure that as the laser power is increased, the broad structures, attributed to $\Delta m_F=1$ coherences in the component graphs, become less and less pronounced. 
However, they are still visible in the difference graphs (Fig.~\ref{fig:im11}, right column), 
although the amplitude slightly decreases, and the sign of the difference signal becomes negative 
for the $\Omega_R=4.50$~MHz (19.6~mW/cm$^2$) case (bottom right in Fig.~\ref{fig:im11}).

Figure~\ref{fig:im12} shows the signal dependence on laser power for the $F_g=1 \longrightarrow F_e=1$ 
transition of the $D_2$ line of $^{87}$Rb. As the magnetic field is increased, after the initial increase of the signal 
due to the dark resonance at 0~G, the signal gradually diminishes.  
However, two small peaks around 45 G and 57 G 
and a broad structure between 7 and 26 G are visible in the component graphs (Fig.~\ref{fig:im12}, left and center columns). 
The structures visible in the graph of the difference signal (Fig.~\ref{fig:im12}, right column) 
must be related to $\Delta m_F=1$ coherences. 
Indeed, the magnetic sublevels $m_F=0$ and $m_F=+1$ of $F_e=1$ cross at 21 G, giving rise to 
the broad structure from 7 to 26 G (see Fig.~\ref{fig:im6}).\\
\indent The small peak at 57 G is caused by the crossing of $m_F=0$ of $F_e=1$ and 
$m_F=-2$ of $F_e=3$ (see Fig.~\ref{fig:im5}), which allows $\Delta m_F=2$ coherences to be created. 
As a result, one can observe a small rise in the component LIF signals.
This peak should vanish as the difference of the components is taken, since it is related to a $\Delta m_F=2$ coherence. In the calculated curve it indeed vanishes, but remains in the measured curve. Possible explanations could be higher-order non-linear effects not treated by the model or even small experimental imperfections.
 
The small peak at 45 G in the component graphs cannot be attributed to any crossing in the excited or the ground states.
The fact that it is visible in the difference graphs might suggest that it is connected to a $\Delta m_F=1$ coherence. 
However, theoretical calculations show that when the Zeeman coherences in the density matrix are ``turned off'' this peak remains, which suggests that it is not connected to any coherences.
While the precise origin of the peak remains unknown, the appearance of this peak in both theory and experiment explicitly shows two things: (i) how nonlinear these signals are and (ii) how well the theoretical model works in describing them.

For each value of the laser power density,  
the theoretical curve which best described the results of the experiment was selected. 
Figure~\ref{fig:im13} shows that the choices made to achieve the best agreement were not arbitrary, but resulted in values that obey the expected relationship between laser power density and Rabi frequency. The laser power density is plotted against the square of the Rabi frequency for which the best fit of the calculated curve to experimental measurements was obtained. 
The points should lie on a straight line, and indeed, they all fall close to the best-fit line. We may conclude that, at least up to these intensity values, the reduced Rabi frequency $\Omega_R$ is proportional to the square root of the intensity $I$. 

\begin{figure}
	\includegraphics[page=1,width=\linewidth]{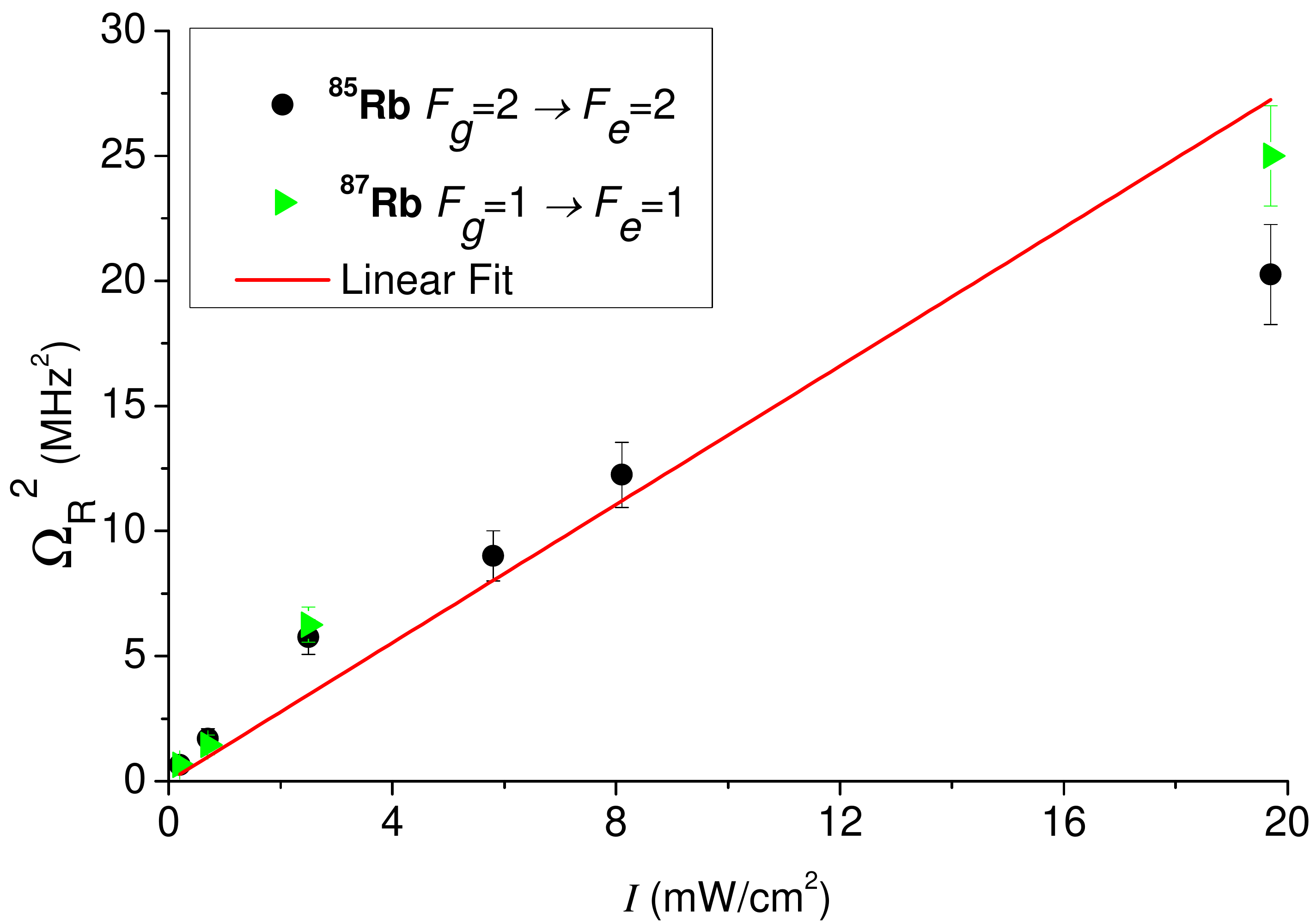}
    \caption{\label{fig:im13}(Color online) Dependence of the squared Rabi frequency $\Omega_R^2$ on the laser power density $I$ 
    together with a linear fit for the transition $F_g=2 \longrightarrow F_e=2$ in $^{85}$Rb 
    and the transition $F_g=1 \longrightarrow F_e=1$ in $^{87}$Rb.}
\end{figure}
\section{\label{sec:level5}Conclusion}
We have carried out experiments with laser power densities that fulfill the nonlinear absorption conditions and developed a theoretical model that describes AOC in these conditions.
The increased magnetic fields and the detection of individual circularly polarized light components in the experiments let us see the structure of the signal in more detail than before~\cite{Alnis:2001}.
With one small exception in Fig.~\ref{fig:im12}, all details, even very small ones, predicted by the theory were reproduced by the experiment 
and were shown to be related to features of the level-crossing diagrams. 
Their positions and relative amplitudes match satisfactorily. 
The signal dependence on laser power density shows that as the laser power increases the structures associated with 
$\Delta m_F=1$ become less pronounced in the individual component signals and the difference signal. 
The signals do not show any visible dependence on the the precise hyperfine transition that is excited from a single ground-state 
hyperfine level.
If the Zeeman splitting of an unknown atom or molecule are of interest, then the measurements of the circularity degree will clearly show whether the splitting is linear or nonlinear, because the circularity degree is nonzero only when the magnetic splitting of Zeeman sublevels is nonlinear, and peaks in this signal will correspond to the crossings of magnetic sublevels. 
The level crossings are determined by the magnetic field value and two constants: magnetic moment and the hyperfine splitting constant. The analysis of level-crossing signals can help to determine these two constants for unknown atomic or molecular systems.

\begin{acknowledgments}
We thank the Latvian State Research Programme (VPP) project IMIS$^2$ and the NATO Science for Peace and Security Programme project SfP983932 ``Novel magnetic sensors and techniques for security applications'' for financial support.

\end{acknowledgments}

\bibliography{AOCBiblioteka}
\end{document}